# Reactive solute transport in physically and chemically heterogeneous porous media with multimodal reactive mineral facies: The Lagrangian approach

Mohamad Reza Soltanian[1,*] Robert Ritzi[1], Zhenxue Dai[2], Chaocheng Huang[3]

[1]Department of Earth and Environmental Sciences, Wright State University, Dayton, OH, 45435

[2]EES-16, Earth and Environmental Sciences Division, Los Alamos National Laboratory, Mailstop T003, Los Alamos, NM 87545

[3]Department of Mathematics and Statistics, Wright State University, Dayton, OH, 45435

*Corresponding author. Phone: (937) 775-2201, e-mail: m.rezasoltanian@gmail.com

## Abstract

Physical and chemical heterogeneities have a large impact on reactive transport in porous media. Examples of heterogeneous attributes affecting reactive mass transport are the hydraulic conductivity ($K$), and the equilibrium sorption distribution coefficient ($K_d$). This paper uses the Deng et al. (2013) conceptual model for multimodal reactive mineral facies and a Lagrangian-based stochastic theory in order to analyze the reactive solute dispersion in three-dimensional anisotropic heterogeneous porous media with hierarchical organization of reactive minerals. An example based on real field data is used to illustrate the time evolution trends of reactive solute dispersion. The results show that the correlation between the hydraulic conductivity and the equilibrium sorption distribution coefficient does have a significant effect on reactive solute dispersion. The anisotropy ratio does not have a significant effect on reactive solute dispersion.

Furthermore, through a sensitivity analysis we investigate the impact of changing the mean, variance, and integral scale of $K$ and $K_d$ on reactive solute dispersion.

**Introduction**

Reactive transport in porous formations is controlled by heterogeneity in physical and chemical properties (Dagan, 1989; Bellin et al., 1993; Miralles-Wilhelm and Gelhar, 1996; Brusseau and Srivastava, 1997; Rajaram, 1997; Dai et al., 2009; Deng et al., 2010; Soltanian et al., 2014). Examples of these properties are hydraulic conductivity ($K$), and the equilibrium sorption distribution coefficient ($K_d$). It has been shown that these parameters are scale-dependent (Allen-King et al., 1998, 2006; Davis et al., 2004; Ritzi et al., 2004; Dai et al., 2007; Ramanathan et al., 2010; Zhang et al., 2013; Ritzi et al., 2013). The spatial variations of physical and chemical heterogeneity are known to be responsible for the scale-dependence of transport parameters such as the retardation factor and the macrodispersivity (Dai et al., 2009; Deng et al., 2013).

Different methods have been proposed for dealing with scale-dependent transport parameters. For example, it is common to use the upscaling process in order to incorporate the effect of small-scale variability on solute transport (Rubin, 2003). Various schemes have been suggested in the literature to upscale reactive transport parameters, as reviewed by Dentz et al. (2011). These include volume averaging (e.g., Whitaker, 1999), stochastic averaging (e.g., Gelhar and Axness, 1983; Dagan, 1984), homogenization (e.g., Lunati et al., 2002), and renormalization (e.g., Zhang, 1998). For example, the time evolution of a conservative solute dispersion has been investigated in detail for unimodal porous media (e.g., Dagan, 1989, Rajaram and Gelhar, 1993; Rubin et al., 1999; Fiori and Dagan, 2002). The time- and scale-dependent effective retardation factor in a unimodal porous media was presented by Rajaram (1997) and in a hierarchical porous media by Deng et al. (2013) using a Lagrangian-based theory. The time evolution of reactive solute dispersion undergoing equilibrium sorption has been investigated by Bellin et al. (1993) and Bellin and Rinaldo (1995) for unimodal porous media. Samper and Yang (2006) analyzed the multicomponent cation exchange reactions in heterogeneous porous media (see also Yang and Samper, 2009). In this paper we use the Deng et al. (2013) conceptual model for porous media with hierarchical organization of reactive minerals (see Fig.1) and develop a Lagrangian-based theory to analyze the reactive solute dispersion undergoing equilibrium sorption.

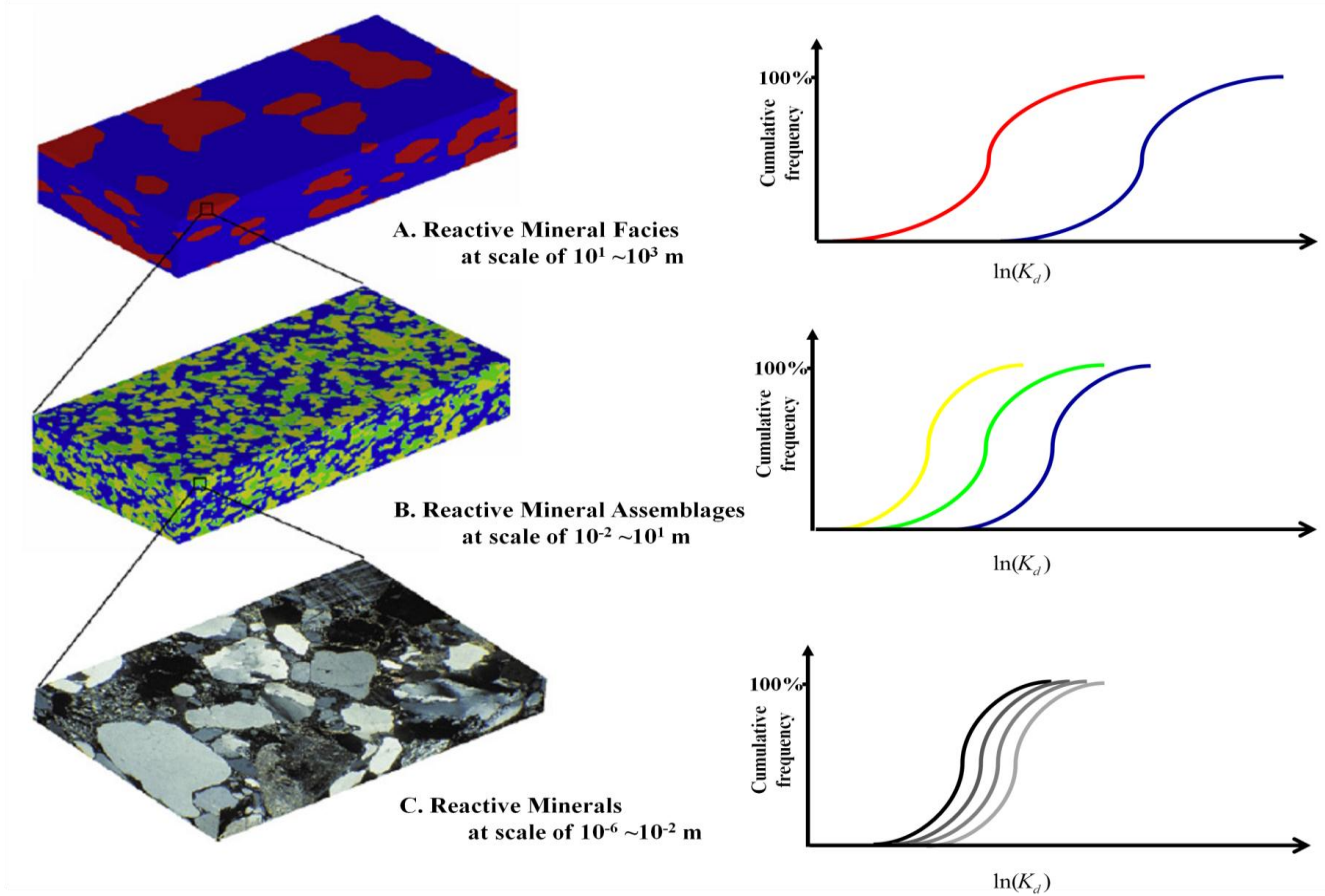

Fig. 1. Conceptual model for reactive mineral facies at different spatial scales, and corresponding modes for the log sorption distribution coefficient, ln ($K_d$). Modified from Deng et al. (2013).

Aquifer architecture is often conceptualize as a hierarchy with facies types defined at each scale comprising assemblages of facies types defined at a smaller scale, across any number of hierarchical levels (e.g. Bridge, 2006). Using the information about facies types defined at different scales can considerably simplify the task of characterizing subsurface heterogeneity. It has been shown that facies types at different scale control the scale-dependence of $K$ and $K_d$ (Allen King et al., 2006; Ritzi and Allen-King, 2007; Dai et al., 2004, 2005; Zhou et al., 2013; Gershenzon et al., 2014). Importantly, $K_d$ is known to vary with sedimentary facies types (Allen-King et al., 1998, 2006; Ritzi et al., 2013; Soltanian and Ritzi, 2014) and reactive mineral facies (Cheng et al., 2007; Deng et al., 2013). Information from sedimentary facies types and/or reactive mineral facies at different scales could be used in developing models for understanding reactive transport processes.

Facies classifications are not unique (Dai et al., 2005; Soltanian and Ritzi, 2014). What is important is that the classification should be useful, and usefulness depends upon context. Reactive minerals can be used to define facies for characterizing heterogeneity both in $K$ and in $K_d$. The Deng et al. (2013) conceptual model is useful for geologic architecture within bedrock in which the type of reactive minerals and their spatial distributions exert the strongest control on the attributes of interest. Using this classification might not be appropriate for deposits where $K$ does not co-vary with mineralogic facies controlling reactivity. We accept that the Deng et al. (2013) classification is useful for geologic architecture within certain settings and our goal is to derive a Lagrangian-based theory using their classification for hierarchical organization of reactive minerals. To our knowledge, there is no theory available for analyzing reactive solute dispersion in hierarchical porous media with hierarchical organization of reactive minerals. Note that the theory developed in this paper can be extended to any type of system classification. In developing the theory we follow Rajaram (1997) which is different in part from the Lagrangian-based model presented by Bellin et al. (1993). Here the theory is developed in a formation with multiple $K$ and $K_d$ modes and hierarchical organization across scales. Note that while Bellin et al. (1993) represented three-dimensional isotropic formations with one scale of spatial variability, in this study we derived a model for three-dimensional anisotropic formations.

In section 2 we briefly review the Deng et al. (2013) conceptual model for hierarchical multimodal porous media with reactive minerals. Section 3 presents the derivation of a Lagrangian-based theory for reactive solute dispersion. In section 4 an example is used to illustrate the utility of the developed theory.

**Conceptual model for hierarchical multimodal porous media and geostatistical characterization**

Aside from aqueous-phase chemical species and physiochemical conditions such as temperature and pH, the sorption reactions in porous media depends on types of reactive minerals and their spatial distributions (Deng et al., 2010, 2013). Mineral reactivity is defined in terms of sorption/desorption process. Deng et al. (2013) presented a conceptual model of reactive minerals in quartz-feldspar sandstone with multimodal $\ln(K_d)$ and $\ln(K)$ subpopulations. In their conceptual model mineral facies have a hierarchical organization and there is a corresponding hierarchy of $\ln(K_d)$ subpopulations (see Fig.1). In the hierarchy, the reactive minerals (RMs) constitute reactive mineral assemblages (RMAs), which, in turn, form reactive mineral facies (RMFs). Here we briefly review the hierarchy of reactive minerals.

The base of the hierarchical organization of reactive minerals is the microform scale ($10^{-6}$ to $10^{-2}$ m). The microform scale is associated with RMs. This scale is related to the scale of mineral grains in a rock. RMs are minerals that are sensitive to a specified geochemical reaction. There is $\ln(K_d)$ subpopulations in each RM. Examples of RMs are calcite, smectite, and hematite which have different sorption coefficients (see also Zavarin et al., 2004). There are also non-reactive minerals (NRMs) with a large volume proportions in quartz-feldspar sandstone. These NRMs (e.g., quartz and feldspar) have low sorption capabilities. The second hierarchical level is the mesoform scale ($10^{-2}$ to $10^{1}$ m). The mesoform scale has RMAs with occurrences of both NRMs and RMs. Examples of RMAs are Clay-Quartz-feldspar, Clay-$Fe_2O_3$-Quartz-Feldspar, and Clay-Organic Mater-Quartz-Feldspar. The RMA composed of RMs has a multimodal structure for uranium sorption coefficients. Also, there can be one or several non-reactive mineral assemblages (NRMAs). The third hierarchical level is the macroform scale ($10^{1}$ to $10^{3}$ m) with reactive mineral facies (RMF). These are a body of rock characterized by an association of RMAs (or RMAs and NRMAs). Two types of RMFs are Calcite-Clay-Organic Matter (CCO-RMF) and Clay-Hematite could be found in sandstone. Similar to Deng et al. (2013), for the purpose of demonstration, only the CCO-RMF with three RMAs is used in this study, and thus flow and transport are assumed to occur within a CCO-RMF. The analysis can be easily extended to also include the RM scale. However, it is not in the scope of this study.

Consider a domain $\Omega$ filled with $N$ number of RMA of mutually exclusive occurrences. Let $Y(x)$ be multimodal spatial random variables for $\ln(K)$ or $\ln(K_d)$ at location $x$. It can be expressed using indicator geostatistics as:

$$Y(x) = \sum_{j=1}^{N} I_j(x) Y_j(x) \qquad (1)$$

where $I_j(x)$ is indictor variable within the domain $\Omega$ and $Y_j(x)$ are variables of the $j$-th RMA. Following Ritzi et al. (2004), the composite mean $M_Y$ and variance $\sigma_Y^2$ of $Y_j(x)$ are computed as (see also Huang and Dai, 2008):

$$M_Y = \sum_{j=1}^{N} p_j m_j \qquad (2)$$

$$\sigma_Y^2 = \sum_{j=1}^{N} p_j \sigma_j^2 + \frac{1}{2}\sum_{i=1}^{N}\sum_{j=1}^{N} p_j p_i (m_i - m_j)^2 \qquad (3)$$

where $p_j$, $m_j$, and $\sigma_j^2$ are volumetric proportion, mean, and variance, respectively. The multimodal covariance function of $\ln(K)$ and $\ln(K_d)$ has been presented from previous studies (see also Dai et al., 2004; and Soltanian et al., 2014):

$$C_Y(\xi) = \sum_{j=1}^{N} p_j^2 \sigma_j^2 e^{-\frac{\xi}{\lambda_j}} + \sum_{i=1}^{N} p_j(1-p_j)\sigma_j^2 e^{-\frac{\xi}{\lambda_\varphi}}$$

$$+ \frac{1}{2}\sum_{i=1}^{N}\sum_{j=1}^{N} p_i p_j (m_i - m_j)^2 e^{-\frac{\xi}{\lambda_I}} \qquad (4)$$

where $\lambda_j$ and $\lambda_I$ are the integral scale of the $j$-th RMA unit and the indicator integral scale of the RMAs, respectively; $\lambda_\varphi = \lambda_j \lambda_I / (\lambda_j + \lambda_I)$.

**The Lagrangian-based theory**

Spatial variability of velocity experienced by reactive solutes is the first step in characterizing reactive solute spreading (Bellin et al., 1993; Rajaram, 1997). The Lagrangian velocity for reactive solutes is as follows:

$$u(x) = \frac{v(x)}{R(x)} \qquad (5)$$

where $u$ is the reactive solute velocity, $v$ is the groundwater velocity, and $R$ is the retardation factor with a constant mean $\bar{R}$, variance $\sigma_R^2$ and a stationary spatial covariance $C_{RR}(\xi)$ (Bellin et al., 1993; Rajaram, 1997). For transport of a reactive solute with the linear equilibrium sorption assumption, spatial variability of $R$ is related to the spatial variability of $K_d$ by the relationship $R(x) = 1 + (\rho_b/n)K_d(x)$ where $\rho_b$ and $n$ are the bulk density and porosity of the medium, respectively. The perturbation of the reactive solute velocity, $u_i'$, is found by linearizing equation (5) as:

$$u_i' = \frac{v_i'}{\bar{R}} - \frac{\bar{v}_i R'}{\bar{R}^2} \qquad (6)$$

where $R'$ is the $R$ perturbation, and $v_i'$ is the perturbation of the groundwater velocity (Rajaram, 1997). Using equation (6), and the spectra of the flow velocity and the retardation factor ($S_{v_1 v_1}$ and $S_{RR}$) and their cross-spectral density, $S_{v_i R}$, the spectral density of the reactive solute velocity is given by:

$$S_{u_i u_j}(k) = \frac{1}{\bar{R}^2} S_{v_i v_j}(k) + \frac{\bar{v}_i \bar{v}_j}{\bar{R}^4} S_{RR}(k) - \frac{\bar{v}_i}{\bar{R}^3} S_{v_j R}(k) - \frac{\bar{v}_j}{\bar{R}^3} S_{v_i R}(k) \qquad (7)$$

where $k = (k_1, k_2, k_3)^T$ is a 3-D wave-number vector (Rajaram, 1997). Thus, the following relationship defines the covariance of reactive solute velocity:

$$C_{u_i u_j}(\xi) = \frac{1}{\bar{R}^2} C_{v_i v_j}(\xi) + \frac{\bar{v}_i \bar{v}_j}{\bar{R}^4} C_{RR}(\xi) - \frac{\bar{v}_i}{\bar{R}^3} C_{v_j R}(\xi) - \frac{\bar{v}_j}{\bar{R}^3} C_{v_i R}(\xi) \qquad (8)$$

Following Dagan (1984, 1989) and Bellin et al. (1993) and using the Lagrangian-based theory the time-dependent dispersion tensor for a reactive solute, $\tilde{D}_{ij}^R(t)$, is found as:

$$\tilde{D}_{ij}^{R}(t) = \int_{0}^{t} C_{u_i u_j}(s)\,ds \qquad (9)$$

We assume that the velocity covariance depends on the mean particle trajectory instead of the actual one. Therefore, $\xi$ can be approximated by $\bar{v}_1 t / \bar{R}$. This assumption has been successfully used for analyzing the dispersion of nonreactive and reactive solutes (Dagan, 1989; Bellin et al., 1993; Bellin and Rinaldo, 1995; Dai et al., 2004).

Bellin et al. (1993) showed that the transverse retarded velocity is independent of the variability of $R$. Our focus is on analyzing the longitudinal dispersion of a reactive plume as has been observed in field experiments (Roberts et al., 1986; Garabedian et al., 1988). Therefore, we assume that the velocity field is uniform in average, and that the mean velocity is aligned with the $x_1$ axis such that $v = (v_1, 0, 0)$. Thus,

$$\tilde{D}_{11}^{R}(t) = \int_{0}^{t} C_{u_1 u_1}\left(\frac{\bar{v}_1}{\bar{R}} s\right) ds$$

$$= \frac{1}{\bar{R}^2} \int_{0}^{t} C_{v_1 v_1}\left(\frac{\bar{v}_1}{\bar{R}} s\right) ds \qquad (10\,\mathrm{a})$$

$$+ \frac{\bar{v}_1^2}{\bar{R}^4} \int_{0}^{t} C_{RR}\left(\frac{\bar{v}_1}{\bar{R}} s\right) ds \qquad (10\,\mathrm{b})$$

$$- \frac{2\bar{v}_1}{\bar{R}^3} \int_{0}^{t} C_{v_1 R}\left(\frac{v_1}{\bar{R}} s\right) ds \qquad (10\mathrm{c})$$

Bellin et al. (1993) used both the cross-covariance of ln($K$) and $R$, and the cross-covariance of the hydraulic head and $R$ by performing first-order approximation in order to derive $C_{v_1 R}(\xi)$. Here we follow Rajaram (1997) in which $C_{v_1 R}(\xi)$ is derived directly from its spectral density, $S_{v_1 R}(k)$, using the cross-correlation between ln($K$) and ln($K_d$).

The retardation factor can also be expressed as $R(x) = 1 + (\rho_b/n)e^{w(x)}$ where $w$ is $\ln(K_d)$. Using stochastic theory $w$ can be replaced by $\bar{w} + w'$ where $\bar{w}$ and $w'$ are the mean and the perturbation of $w$. Soltanian et al. (in revision) derived the following expression for the perturbation of $R$:

$$R' = R - \bar{R} = \frac{\rho_b}{n} K_d^G (e^{w'} - e^{[\frac{\sigma_w^2}{2}]}) \qquad (11)$$

where $K_d^G$ is the geometric mean of $K_d(x)$, and $\sigma_w^2$ is the variance of $w$. Using the Taylor series expansion for $e^{w'}$, we calculate $C_{RR}$ approximately as:

$$C_{RR}(\xi) = (\frac{\rho_b}{n} K_d^G)^2 e^{[\sigma_w^2]} (e^{C_w(\xi)} - 1) \qquad (12)$$

Therefore, the corresponding variance of the retardation factor is:

$$\sigma_R^2 = (\frac{\rho_b}{n} K_d^G)^2 e^{[\sigma_w^2]} (e^{[\sigma_w^2]} - 1) \qquad (13)$$

Following Gelhar (1993), Rajaram (1997), and Deng (2013) the perturbation of $v$ is obtained as:

$$v' = \frac{K^G J}{n} (1 - \frac{k_1^2}{k^2}) f \qquad (14)$$

where $f$ is the perturbation of $\ln(K)$.

Note that we used a nonlinear expansion for $\ln(K_d)$, equation (11), and first-order for $\ln(K)$, equation (14). Bellin et al. (1993) and Bellin and Rinaldo (1995) have used the same inconsistent expansion in order to analyze the time-dependent dispersion of reactive solutes (see equation (10a) and (17) in Bellin et al., 1993). Their results were tested against numerical simulations and validated by Bosma et al. (1993) for relatively small variances (<1.6). Here we assume similarly small variances. Thus, the developed theory is valid for aquifers with mild heterogeneity contrast ($\sigma_f^2$, $\sigma_w^2 <1$).

The $S_{v_1 R}(k)$ is derived by using equations (11) and (14). The $S_{v_1 R}(k)$ is found as:

$$S_{v_1 R}(k) = \frac{\rho_b}{n^2} K^G K_d^G J (1-\frac{k_1^2}{k^2}) f (e^{w'} - e^{[\frac{\sigma_w^2}{2}]}) \tag{15}$$

Using the Taylor series expansion for $e^{w'}$ and considering the point that the odd moments of a log normal distribution are zero, $S_{v_1 R}(k)$ is found as:

$$S_{v_1 R}(k) = \frac{\rho_b}{n^2} K^G K_d^G J (1-\frac{k_1^2}{k^2}) \frac{\sinh(\sigma_w)}{\sigma_w} S_{fw}(k) \tag{16}$$

where $S_{fw}(k)$ is the spectral density of the fluctuations of $f - w$, and sinh is hyperbolic sine function. The $K$ and $K_d$ are assumed to be perfectly correlated as $\ln(K_d)= a \ln(K)+ b$, where $a$ and $b$ are real constants. Then the cross-spectral density of $\ln(K)$ and $\ln(K_d)$ has a linear relationship given by $\ln(K)$, i.e. $S_{fw}(k) = a S_{ff}(k)$. Therefore, $S_{v_1 R}(k)$ is expressed as:

$$S_{v_1 R}(k) = \frac{\rho_b}{n^2} K^G K_d^G J a \frac{\sinh(\sigma_w)}{\sigma_w} (1-\frac{k_1^2}{k^2}) S_{ff}(k) \tag{17}$$

Using equation (14) $S_{v_1 v_1}(k)$ is easily found as follows:

$$S_{v_1 v_1}(k) = (\frac{K^G J}{n})^2 (1-\frac{k_1^2}{k^2})^2 S_{ff}(k) \tag{18}$$

The relationship between the spectral density and the covariance function is expressed as follows:

$$S(k) = \frac{1}{(2\pi)^3} \int\int\int_{-\infty}^{\infty} e^{-ik.\xi} C(\xi) d\xi \qquad (19)$$

The $C_{v_1 v_1}(\xi)$ and $C_{v_1 R}(\xi)$ are used in form of their spectral density to find the longitudinal dispersion of a reactive solute. Therefore, equation (10) can be rewritten as follows:

$$\tilde{D}_{11}^R(t) = \int_0^t C_{u_1 u_1}(\frac{\overline{v}_1}{\overline{R}} s) ds$$

$$= \frac{1}{\overline{R}^2} \int_0^t \int\int\int_\Omega e^{-ik\frac{\overline{v}_1}{\overline{R}}s} S_{v_1 v_1}(k) \, ds \, dk \qquad (20a)$$

$$+ \frac{\overline{v}_1^2}{\overline{R}^4} \int_0^t C_{RR}(\frac{\overline{v}_1}{\overline{R}} s) ds \qquad (20b)$$

$$- \frac{2\overline{v}_1}{\overline{R}^3} \int_0^t \int\int\int_\Omega e^{-ik\frac{\overline{v}_1}{\overline{R}}s} S_{v_1 R}(k) \, ds \, dk \qquad (20c)$$

By substituting equations (12), (17), and (18) into equation (20) and using equation (4) to represent the multimodal covariance function of $\ln(K)$ and $\ln(K_d)$, the final expression for longitudinal dispersivity of a reactive solute in multimodal porous media is found as:

$$\alpha_{11}^R(t) = \frac{\tilde{D}_{11}^R(t) \overline{R}}{\overline{v}_1} = \sum_{j=1}^{N} p_j^2 \sigma_{fi}^2 \lambda_i F_1(\lambda_i) + \sum_{j=1}^{N} p_j(1-p_j) \sigma_{fi}^2 \lambda_\varphi F_1(\lambda_\varphi) + \frac{1}{2} \sum_{i=1}^{N}\sum_{j=1}^{N} p_i p_j (m_{fi} - m_{fj})^2 \lambda_I F_1(\lambda_I)$$

$$+ \frac{\overline{v}_1}{\overline{R}^3} (\frac{\rho_b}{n} K_d^G)^2 e^{[\sigma_w^2]} (\int_0^t (e^{\sum_{j=1}^{N} p_j^2 \sigma_{wj}^2 e^{-\frac{\overline{v}_1}{\overline{R}\lambda_i}s} + \sum_{i=1}^{N} p_j(1-p_j)\sigma_{wj}^2 e^{-\frac{\overline{v}_1}{\overline{R}\lambda_\varphi}s} + \frac{1}{2}\sum_{i=1}^{N}\sum_{j=1}^{N} p_i p_j (m_{wi} - m_{wj})^2 e^{-\frac{\overline{v}_1}{\overline{R}\lambda_I}s}} - 1) ds)$$

$$- \frac{2}{\overline{R}} \frac{\rho_b}{n^2} K^G K_d^G J a \frac{\sinh(\sigma_w)}{\sigma_w} \{\sum_{j=1}^{N} p_j^2 \sigma_{fi}^2 \lambda_i F_2(\lambda_i) + \sum_{j=1}^{N} p_j(1-p_j) \sigma_{fi}^2 \lambda_\varphi F_2(\lambda_\varphi)$$

$$+ \frac{1}{2} \sum_{i=1}^{N}\sum_{j=1}^{N} p_i p_j (m_{fi} - m_{fj})^2 \lambda_I F_2(\lambda_I)\} \qquad (21)$$

where

$$F_1(\lambda) = \{1 - e^{-\tau} - \varepsilon \int_0^\infty 2r J_1(\beta) \frac{2u^{\frac{3}{2}} - \varepsilon r(v + 2u)}{v^2 u^{\frac{3}{2}}}$$

$$+ [\frac{(2-\beta^2)J_1(\beta) - \beta J_0(\beta)}{r\tau^2}][\frac{\varepsilon^3 r^3(v+4u) + u^{\frac{3}{2}}(5v-4u)}{v^3 u^{\frac{3}{2}}}]dr\},$$

$$F_2(\lambda) = \{1 - e^{-\tau} - 2\varepsilon^2 \int_0^\infty r^2 [\frac{1}{2vu^{\frac{3}{2}}} + \frac{1}{v^2 u^{\frac{1}{2}}} - \frac{1}{v^2 \varepsilon r}] J_1(\beta) \, dr\},$$

$$\beta = r\tau,$$

$$u = 1 + r^2,$$

$$v = 1 + r^2 - \varepsilon^2 r^2,$$

$$\tau = \frac{\bar{v}_1 t}{R\lambda},$$

Also, $\varepsilon$ is the anisotropy ratio is defined as the vertical integral scale of the hydraulic conductivity to the horizontal component. The $J_0$ and $J_1$ are the zero and first order Bessel functions, respectively. The integration method to find equation (21) can be found in Appendix A, B, C, and D. The expressions in equation (21) cannot be integrated in closed form. We use the quadrature method in order to numerically integrate and evaluate equation (21) at a number of points in the parameter space, and present the results in Section 4.

**Results and Discussion**

In order to analyze the reactive solute dispersion using the developed Lagrangian-based model, we use an example presented by Deng et al. (2013) as summarized in Table 1. In this example the parameter values of the three RMAs are extracted from a real case.

Table 1. Parameters from the example used by Deng et al. (2013).

| RMF | RMA | $L_j$ | $p_j$ | Parameters | $m_j$ | $\sigma_j^2$ | $M_j^G$ | $\lambda_j$ | $\lambda_\varphi$ | $R_j$ |
|---|---|---|---|---|---|---|---|---|---|---|
| Cc-Clay-OM RMF | Cc-QF | 50.0 | 0.6 | $\ln K$ | 1.5 | 0.6 | 4.48 | 10 | 6.67 | 2.39 |
|  |  |  |  | $\ln K_d$ | -2.2 | 0.22 | 0.11 | 12 | 7.5 |  |
|  | Clay-Fe$_2$O$_3$-QF | 23.5 | 0.15 | $\ln K$ | 0.5 | 0.3 | 1.65 | 6 | 4.62 | 4.76 |
|  |  |  |  | $\ln K_d$ | -1.2 | 0.12 | 0.3 | 8 | 5.71 |  |
|  | Clay-OM-QF | 26.7 | 0.25 | $\ln K$ | 0.05 | 0.15 | 1.05 | 9 | 6.21 | 10.26 |
|  |  |  |  | $\ln K_d$ | -0.3 | 0.1 | 0.74 | 7 | 5.19 |  |

| Parameters | $\ln K$ | | | $\ln K_d$ | | | $R$ | | |
|---|---|---|---|---|---|---|---|---|---|
| Statistics | $M_Y$ | $\sigma_Y^2$ | $M_Y^G$ | $M_Y$ | $\sigma_Y^2$ | $M_Y^G$ | $M_Y$ | $\sigma_Y^2$ | $M_Y^G$ |
| Values | 0.99 | 0.86 | 2.68 | -1.58 | 0.84 | 0.21 | 4.93 | 20.25 | 3.59 |

| Parameters | $\bar{v}_1$ | $\sigma_{v_1R}^2$ | $\lambda_I$ | $n$ | $\rho_b$ | $J$ |
|---|---|---|---|---|---|---|
| Values | 0.21 | 0.228 | 20.0 | 0.2 | 2.5 | 0.01 |

Note: RMA = reactive mineral assemblage within a reactive mineral facies (RMF), Cc = calcite, Fe$_2$O$_3$ = iron oxides, QF = quartz and feldspar; For $j$-th RMA ($j$ = 1, 2, 3), $L_j$ = facies mean length (m), $M_j^G$ = geometric mean, $\lambda_j$ = correlation length (m), $\lambda_\varphi = \lambda_i\lambda_I/(\lambda_i+\lambda_I)$, $R_j$ = retardation factor; $M_Y$ = global mean, $\sigma_Y^2$ = global variance, $M_Y^G$ = global geometric mean, $\bar{v}_1$ = mean flow velocity (m/d), $\sigma_{v_1R}^2$ = cross-covariance of flow velocity and retardation factor, $\lambda_I$ = indicator correlation length (m), $n$ = porosity, $\rho_b$ = bulk density of the porous media (g/cm$^3$), $J$ = average hydraulic gradient, $K$ = hydraulic conductivity (m/d), $K_d$ = sorption coefficient (cm$^3$/g).

The $\alpha_{11}^R(t)$ is plotted in Fig.2 for three cases of correlation between ln($K$) and ln($K_d$): positively correlated ($a$ = 1), uncorrelated ($a$ = 0), and negatively correlated ($a$ = -1). The $\alpha_{11}^R(t)$ increases monotonically with time for all cases. Fig.2 shows that the values of $\alpha_{11}^R(t)$ for the negatively correlated case are larger than those of the uncorrelated and positively correlated cases. In all cases the value of $\alpha_{11}^R(t)$ converges to a constant value when time is sufficiently large. Fig.2 also shows the influence of anisotropy ratio, $\varepsilon$, on $\alpha_{11}^R(t)$. It is observed that $\varepsilon$ has a relatively small impact upon dispersion of reactive solutes in this example. The $\alpha_{11}^R(t)$ is slightly enhanced at early times by a smaller $\varepsilon$ due to lateral mass transfer between streamlines with different velocities located adjacent to each other (Rubin, 2003). Furthermore, as shown in section 4, heterogeneity in ln($K_d$) has a relatively larger impact on reactive solute transport than ln($K$) heterogeneity or its anisotropy ratio.

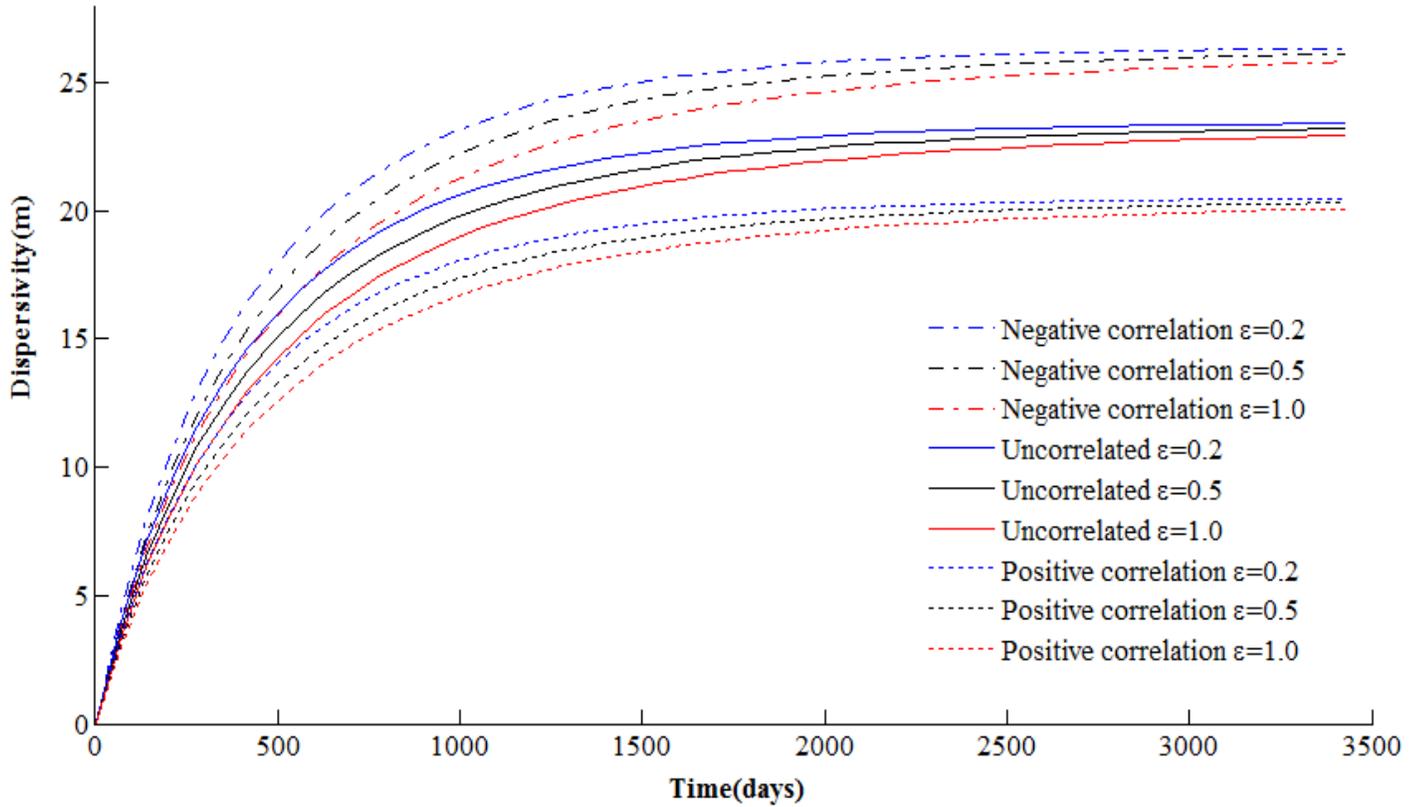

Fig. 2. The longitudinal dispersivity $\alpha_{11}^R(t)$ for a porous medium with three RMFs. Three different cases with positive correlation (a=1), no correlation (a=0), and negative correlation (a=-1) are shown. The influence of the anisotropy ratio, $\varepsilon$, for all of the three cases is presented.

Fig. 3 shows $\alpha_{11}^R(t)$ changes with the indicator correlation length ($\lambda_I$) when the time is fixed at 1000 d. In all cases $\alpha_{11}^R(t)$ increases to a maximum at about $\lambda_I$ =300 m, and then it stays constant. Although $\alpha_{11}^R(t)$ reaches a maximum for three cases, it reaches different values. This reflects the contribution from the cross-correlation between $R$ and $v_1$, represented by equation (20c).

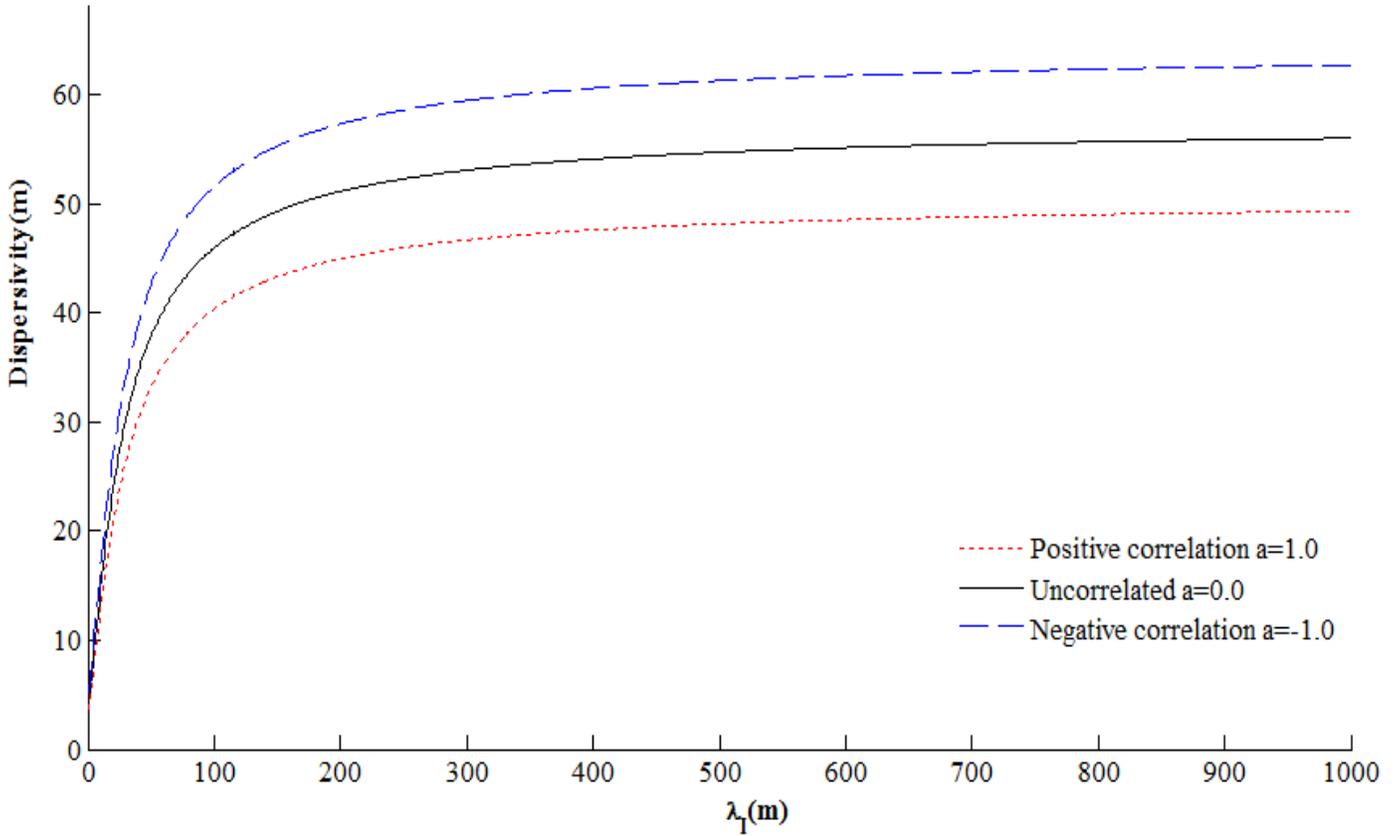

Fig. 3. The longitudinal dispersivity $\alpha_{11}^{R}(t)$ changes with indicator correlation length when time is fixed at 1000 d. Three different cases with positive correlation (a=1), no correlation (a=0), and negative correlation (a=-1) are shown. The anisotropy ratio, $\varepsilon$, is 1.

In order to better understand the impact of heterogeneity in both ln(*K*) and ln($K_d$) on reactive solute dispersion we study the effect of changing the mean, variance, and integral scale of ln(*K*) and ln($K_d$) for each RMA below.

**The impact of ln(*K*) heterogeneity**

The influence of changing the mean of log hydraulic conductivity ($m_j$) for each RMA is shown in Fig. 4A. We set $m_j$ between -2.5 and 2.5 and time is fixed at 1000 d. Fig. 4A shows that any changes in values of mean of hydraulic conductivity can affect $\alpha_{11}^{R}(t)$. However, it depends on the type of RMA. For example, decreasing the mean hydraulic conductivity for RMA1 does not significantly affect $\alpha_{11}^R(t)$ for both positive and negative correlation, whereas $\alpha_{11}^R(t)$ can increase if the mean of hydraulic conductivity for RMA 1 increases. For RMA 2, by decreasing the mean of hydraulic conductivity $\alpha_{11}^R(t)$ increases. The same is true for RMA 3. Note that RMA 1 has the largest variance and mean of hydraulic conductivity and the smallest mean sorption. In contrast, RMA 3 has the smallest variance and mean of hydraulic conductivity and the largest mean sorption coefficient.

Fig. 4B illustrates how changes in variances of log hydraulic conductivity for RMAs can affect $\alpha_{11}^R(t)$ when the time is fixed at 1000 d. Increases in variance of hydraulic conductivity for all RMAs lead to a linear increase in $\alpha_{11}^R(t)$. The dispersivity is more sensitive to changes in variance of log hydraulic conductivity for RMA1 than other two RMAs which can be attributed to the fact that RMA1 has the largest volume proportion. The $\alpha_{11}^R(t)$ is more sensitive to changes in variance of log hydraulic conductivity for RMA 3 than RMA 2 because the volume proportion of RMA 3 is slightly larger than RMA 2.

Fig. 4C shows $\alpha_{11}^R(t)$ changes with integral scales of log hydraulic conductivity when time is fixed at 1000 d. The changes in integral scales of log hydraulic conductivity for RMA 2 and 3 only produce a very small change in $\alpha_{11}^R(t)$. In contrast, for RMA1 it produces a large change in $\alpha_{11}^R(t)$ with a maximum at about $\lambda_j$ =300 m, and then it remains constant. Therefore, $\alpha_{11}^R(t)$ is more sensitive to the integral scale of log hydraulic conductivity for RMA 1 because it has the largest volume proportion and also the largest integral scale.

**The impact of ln($K_d$) heterogeneity**

Fig. 4D shows how $\alpha_{11}^R(t)$ changes with the mean log sorption distribution coefficient when time is fixed at 1000 d. In general, increases from -2.5 to 0 result in slight decreases in $\alpha_{11}^R(t)$. However, increases from 0 to 2.5 lead to increases $\alpha_{11}^R(t)$, more for RMA 2 and 3.

Fig. 4E illustrates how the changes in the variance of the log sorption distribution coefficient for RMAs affect $\alpha_{11}^R(t)$ when the time is fixed at 1000 d. Increases in variances of the sorption distribution coefficient for all RMAs lead to increases in $\alpha_{11}^R(t)$. In contrast to changing the variance of ln($K$), increasing the variance of ln($K_d$) leads to nonlinear increases in $\alpha_{11}^R(t)$. By comparing Fig. 4E to Fig. 4B it is clear that $\alpha_{11}^R(t)$ is more sensitive to changes in the variance of the log sorption distribution coefficient for RMAs than the variance of the log hydraulic conductivity for RMAs.

Fig. 4F shows $\alpha_{11}^R(t)$ changes with the integral scales of the log sorption distribution coefficient when time is fixed at 1000 d. The $\alpha_{11}^R(t)$ is sensitive to changes in the integral scale of the log sorption distribution coefficient for RMA 1 which has the largest volume proportion and also the largest integral scale.

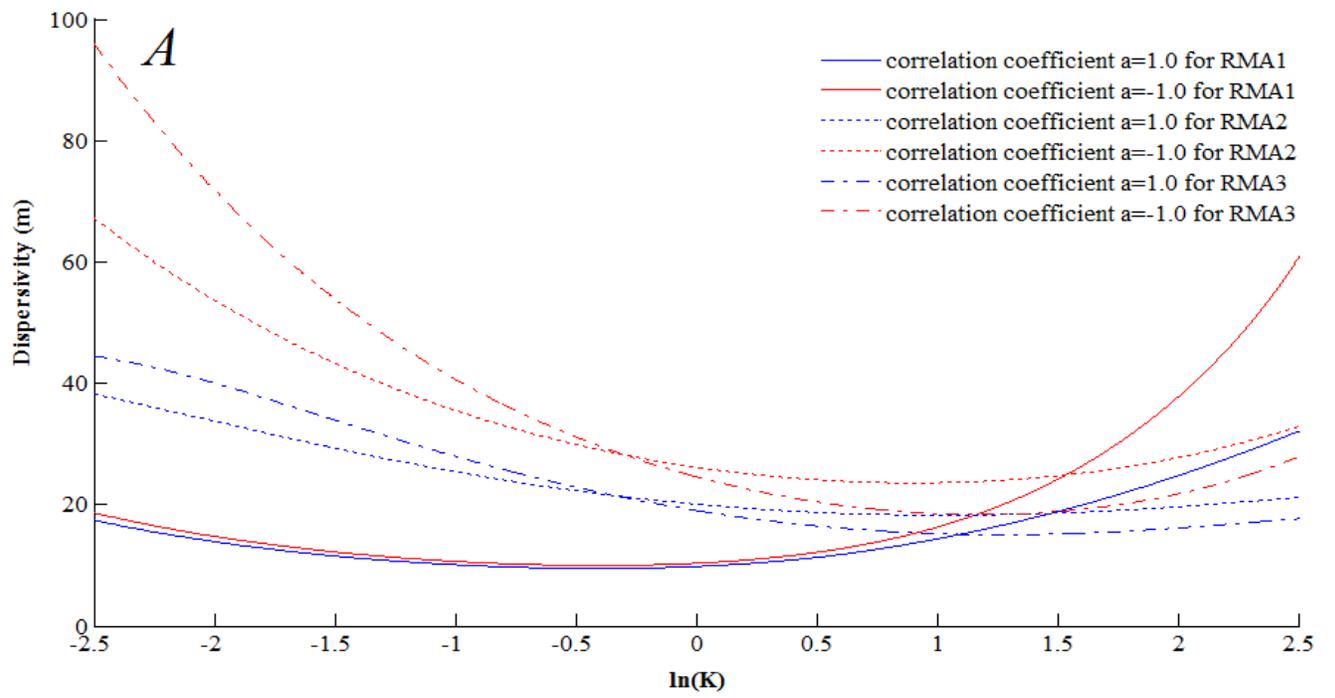
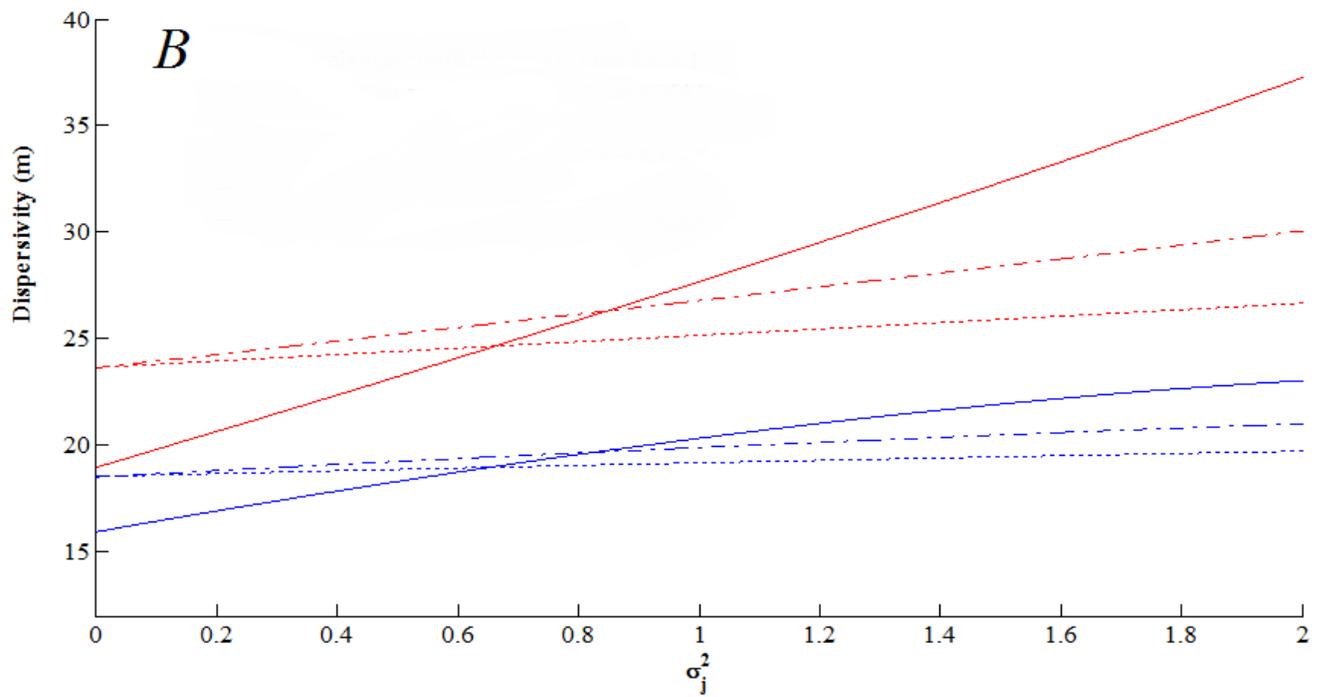

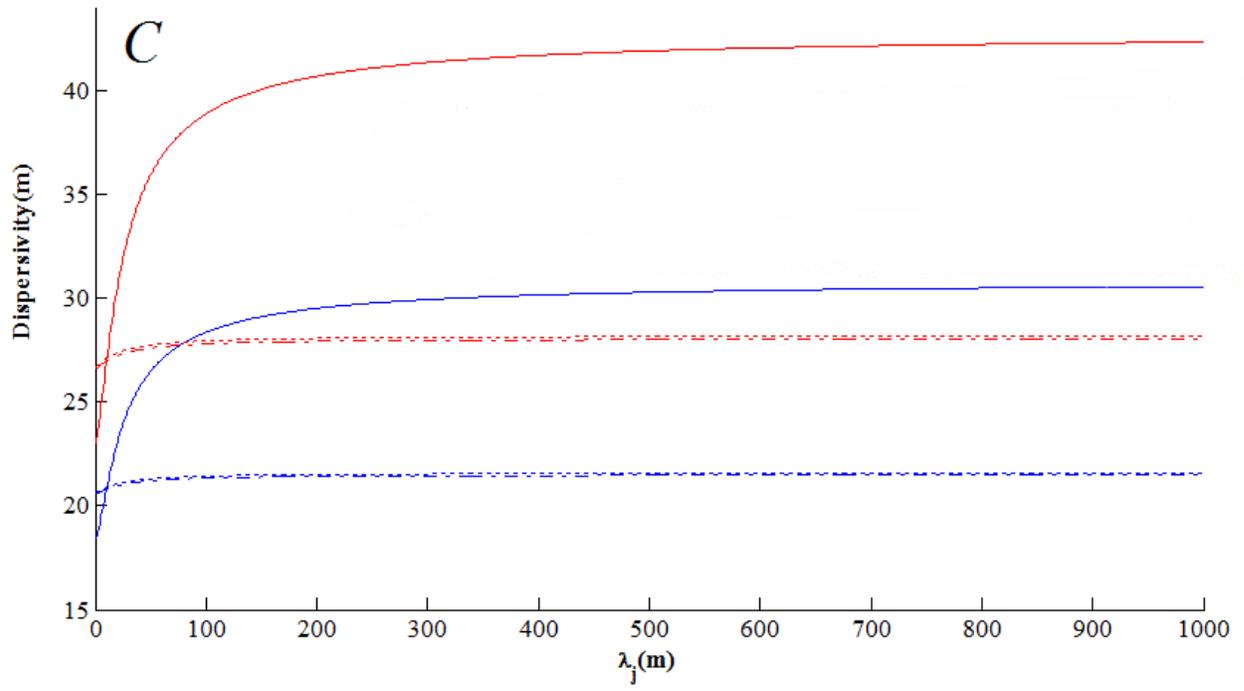

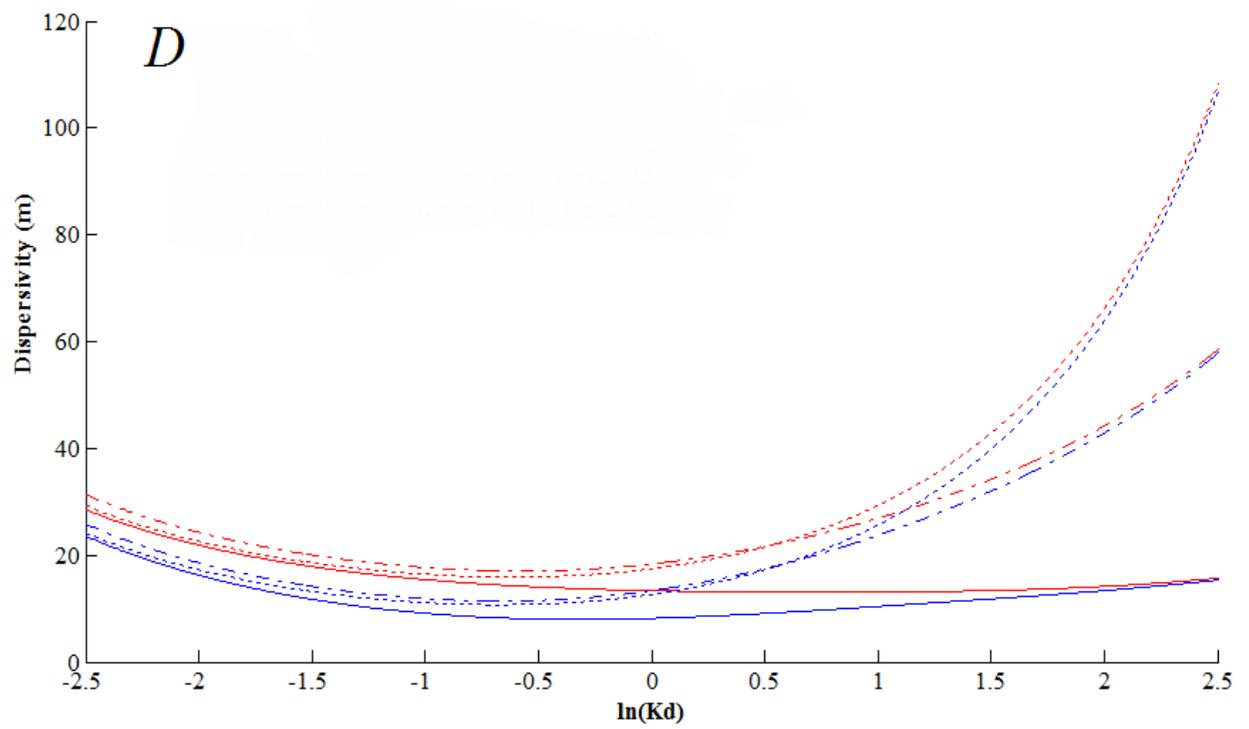

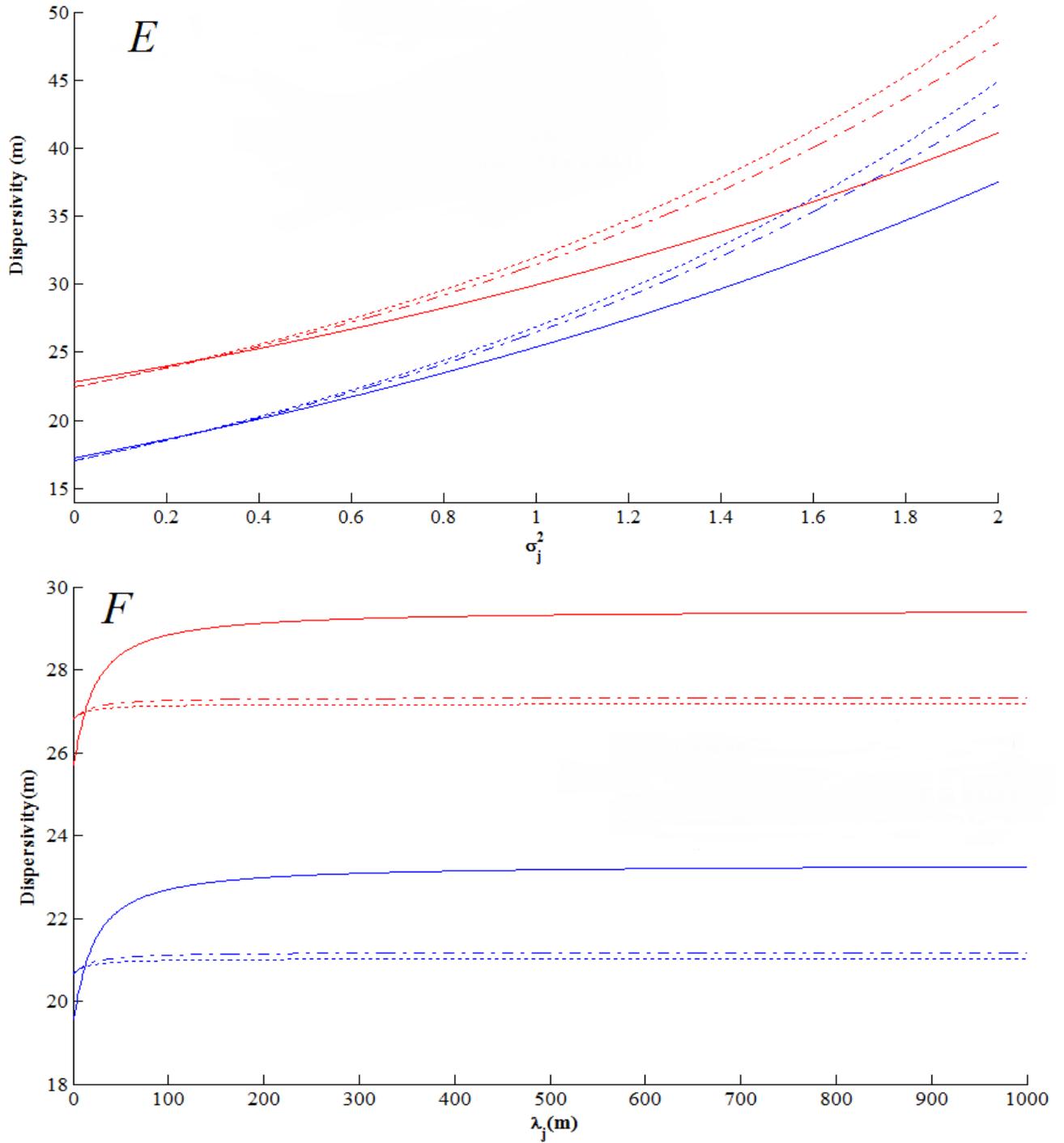

Fig. 4. The longitudinal dispersivity $\alpha_{11}^{R}(t)$ changes with (A) the mean hydraulic conductivity; (B) the variance of the hydraulic conductivity; (C) the integral scale of hydraulic conductivity; (D) the mean of the sorption distribution coefficient; (E) the variance of the sorption distribution coefficient; (F) the integral scale of the sorption distribution coefficient. Two different cases with positive correlation (*a*=1) and negative correlation (*a*=-1) are shown for each RMA. Time at 1000 d. Anisotropy ratio, $\varepsilon$, is 1.

## Conclusions

We developed a Lagrangian-based theory for analyzing the time-dependent reactive solute dispersion in hierarchical porous media with multimodal reactive mineral facies. This study demonstrates that the cross-correlation between *K* and $K_d$ has a large impact on the longitudinal dispersivity $\alpha_{11}^R(t)$. The scale-dependence of reactive solute dispersion originated from heterogeneity in both *K* and $K_d$ and their cross-correlation.

We considered three types of cross-correlation: perfectly positive cross-correlation, perfectly negative cross-correlation, and no cross-correlation. The positive cross-correlation reduces the reactive solute dispersivity, since heterogeneities in *K* and $K_d$ counteract each other, which means that low *K* regions occurs with low retardation or high K regions occurs with high retardation. The negative cross-correlation causes an opposite effect. The results also indicate that anisotropy ratio does not significantly affect the transport of reactive solutes undergoing equilibrium sorption in the example used.

Furthermore, reactive solute dispersion is scale-dependent but not a linear function of the indicator correlation scale. We also performed analyses of sensitivity to changes in the integral scales of ln(*K*) and ln($K_d$), and in the means and variances of ln(*K*) and ln($K_d$) for each reactive mineral assemblage. The results show that heterogeneity in both ln(*K*) and ln($K_d$) can significantly affect $\alpha_{11}^R(t)$. The $\alpha_{11}^R(t)$ is very sensitive to changes in mean and variance of both $\ln(K)$ and $\ln(K_d)$. The $\alpha_{11}^R(t)$ is most sensitive to changes of integral scales of $\ln(K)$ and $\ln(K_d)$ for the reactive mineral assemblage with a larger volume proportion.

**Appendix A. Derivation of spectral density of fluctuations in ln(K)**

Considering an exponential covariance function of $\ln(K)$:

$$C_{ff}(\xi) = \sigma_f^2 e^{(-|\frac{\xi}{\mu}|)} = \sigma_f^2 e^{(-\sqrt{(\frac{\xi_1}{\mu_1})^2 + (\frac{\xi_2}{\mu_2})^2 + (\frac{\xi_3}{\mu_3})^2})} \qquad (A1)$$

where $\mu_i$ are integral scale. The spectral density of $\ln(K)$ is evaluated by taking the Fourier transform of equation (19). Substituting (A1) into (19), then it yields:

$$S(k) = \frac{1}{(2\pi)^3} \int\int\int_{-\infty}^{\infty} e^{-ik\cdot\xi} \sigma_f^2 e^{(-|\frac{\xi}{\mu}|)} d\xi \qquad (A2)$$

Integrating (A2) leads to the following expression for spectral density of $\ln(K)$:

$$S_{ff}(k) = \frac{\sigma_f^2 \mu_1 \mu_2 \mu_3}{\pi^2} \frac{1}{(1+(\mu k)^2)^2} \qquad (A3)$$

where $(\mu k)^2 = (\mu_1 k_1)^2 + (\mu_2 k_2)^2 + (\mu_3 k_3)^2$. Note that (A3) is also used for spectral density of $\ln(K_d)$.

**Appendix B. Derivation of equation (20a)**

Equation (20 a) contains the following integral:

$$(20a) = \frac{1}{R^2} \int_0^t C_{v_1 v_1}(\frac{\bar{v}_1}{R} s, 0, 0) ds$$

Based on equation (18) we can write the following:

$$S_{v_1 v_1}(k) = (\frac{K^G J}{n})^2 (1-\frac{k_1^2}{k^2})^2 S_{ff}(k) = \bar{v}_1^2 (1-\frac{k_1^2}{k^2})^2 S_{ff}(k) = \frac{\bar{v}_1^2 \sigma_f^2 \mu_1 \mu_2 \mu_3}{\pi^2} (1-\frac{k_1^2}{k^2})^2 \frac{1}{(1+(\mu k)^2)^2}$$

Therefore,

$$C_{v_1 v_1}(\xi) = \frac{\bar{v}_1^2 \sigma_f^2 \mu_1 \mu_2 \mu_3}{\pi^2} \int_{-\infty}^{\infty}\int_{-\infty}^{\infty}\int_{-\infty}^{\infty} (1-(\frac{k_1}{k})^2)^2 \frac{1}{(1+(\mu k)^2)^2} \cos(k.\xi) \, dk$$

Hence,

$$C_{v_1 v_1}(\frac{\bar{v}_1}{R} s, 0, 0) = \frac{\bar{v}_1^2 \sigma_f^2 \mu_1 \mu_2 \mu_3}{\pi^2} \int_{-\infty}^{\infty}\int_{-\infty}^{\infty}\int_{-\infty}^{\infty} (1-(\frac{k_1}{k})^2)^2 \frac{1}{(1+(\mu k)^2)^2} \cos(k_1 \frac{\bar{v}_1}{R} s) \, dk$$

$$\underset{k_i' = \mu_i k_i}{=} \frac{\bar{v}_1^2 \sigma_f^2}{\pi^2} \int_{-\infty}^{\infty}\int_{-\infty}^{\infty}\int_{-\infty}^{\infty} (\frac{\mu_2^{-2} k_2^2 + \mu_3^{-2} k_3^2}{\mu_1^{-2} k_1^2 + \mu_2^{-2} k_2^2 + \mu_3^{-2} k_3^2})^2 \frac{1}{(1+k^2)^2} \cos(k_1 \frac{\bar{v}_1}{R} s) \, dk$$

Changing variables $s' = \mu_1^{-1} \bar{R}^{-1} \bar{v}_1 s$ and $\tau = \bar{v}_1 t / \mu_1 \bar{R}$ gives:

$$(20a) = \frac{\bar{v}_1}{\bar{R}} \frac{\sigma_f^2 \mu_1}{\pi^2} \int_0^\tau \int_{-\infty}^{\infty}\int_{-\infty}^{\infty}\int_{-\infty}^{\infty} (\frac{\mu_2^{-2} k_2^2 + \mu_3^{-2} k_3^2}{\mu_1^{-2} k_1^2 + \mu_2^{-2} k_2^2 + \mu_3^{-2} k_3^2})^2 \frac{1}{(1+k^2)^2} \cos(k_1 s') \, dk \, ds'$$

Now we have $\mu_1 = \mu_2 = \lambda$ and $\mu_3 = \varepsilon \lambda$ which $\varepsilon$ is the anisotropy ratio. Thus,

$$(20a) = \frac{\bar{v}_1}{\bar{R}} \frac{\sigma_f^2 \lambda}{\pi^2} \int_0^\tau \int_{-\infty}^{\infty}\int_{-\infty}^{\infty}\int_{-\infty}^{\infty} (1 - \frac{k_1^2}{k_1^2 + k_2^2 + \varepsilon^{-2} k_3^2})^2 \frac{1}{(1+k^2)^2} \cos(k_1 s) \, dk \, ds$$

Now changing to spherical coordinate system and defining:

$$k_1 = r\cos\beta, \ k_2 = r\sin\beta, \ k_3 = k_3$$
$$dk_1 k_2 k_3 = r \, dr \, d\theta \, dk_3$$

Therefore,

$$(20a) = \frac{\bar{v}_1}{\bar{R}} \frac{\sigma_f^2 \lambda}{\pi^2} \int_0^\tau ds \int_0^\infty r\, dr \int_0^{2\pi} d\theta \int_{-\infty}^\infty (1 - \frac{r^2 \cos^2\theta}{r^2 + \varepsilon^{-2}k_3^2})^2 \frac{\cos(sr\cos\theta)}{(1+r^2+k_3^2)^2} dk_3 \quad \text{(B1)}$$

In order to derive the above integral the following general integrals are used:

$$\int_{-\infty}^\infty \frac{1}{(a^2+x^2)^2} dx = \frac{\pi}{2a^3} \quad \text{(B2a)}$$

$$\int_{-\infty}^\infty \frac{1}{(a^2+x^2)^2(b^2+x^2)} dx = \frac{\pi}{2(b^2-a^2)a^3} - \frac{\pi}{(b^2-a^2)^2}(\frac{1}{a}-\frac{1}{b}) \quad \text{(B2b)}$$

$$\int_{-\infty}^\infty \frac{1}{(a^2+x^2)^2(b^2+x^2)^2} dx = \frac{\pi}{2(b^2-a^2)^2}(\frac{1}{a^3}+\frac{1}{b^3}) - \frac{\pi}{(b^2-a^2)^3}(\frac{1}{a}-\frac{1}{b}) \quad \text{(B2c)}$$

In case of integral in (B1) the following parameters are used in (B2a)-(B2c):

$$a = 1+r^2 = u,$$
$$b = \varepsilon r,$$
$$a^2 - b^2 = 1 + r^2 - \varepsilon^2 r^2,$$

Thus, one can derive (B1) as follows:

$$(20a) = \frac{\bar{v}_1}{\bar{R}} \sigma_f^2 \lambda \int_0^\tau ds \int_0^\infty \{\frac{r}{2u^{\frac{3}{2}}} A(sr)dr + 2\varepsilon^2 r^3 [\frac{1}{2vu^{\frac{3}{2}}} + \frac{1}{v^2 u^{\frac{3}{2}}} - \frac{1}{v^2 \varepsilon r}]B(sr)$$

$$+ \varepsilon^4 r^5 [\frac{1}{2v^2 u^{\frac{3}{2}}} + \frac{1}{2v^2(\varepsilon r)^3} + \frac{2}{v^3 u^{\frac{1}{2}}} - \frac{2}{v^3 \varepsilon r}]C(sr)\}dr$$

where

$$A(sr) = 2J_o(sr) \quad \text{(B3a)}$$

$$B(sr) = \frac{2J_o(sr)sr - 2J_1(sr)}{sr} \quad \text{(B3b)}$$

$$C(sr) = \frac{2(sr)^3 J_o(sr) - 6J_o(sr)(sr) + 12J_1(sr) - 4J_1(sr)(sr)^2}{(sr)^3} \quad \text{(B3c)}$$

The $J_o$ and $J_1$ are the zero and first order Bessel functions, respectively. Now we change the variable $\beta = r\tau$. Thus,

$$(20a) = \frac{1}{\overline{R}}\{\overline{v}_1 \sigma_f^2 \lambda \int_0^\tau ds \int_0^\infty \frac{r}{2u^{\frac{3}{2}}} A(sr) dr$$

$$+ \overline{v}_1 \sigma_f^2 \lambda \int_0^\infty 4\varepsilon^2 r^3 [\frac{1}{2vu^{\frac{3}{2}}} + \frac{1}{v^2 u^{\frac{3}{2}}} - \frac{1}{v^2 \varepsilon r}] J_1(\beta) dr$$

$$+ \overline{v}_1 \sigma_f^2 \lambda \int_0^\infty \varepsilon^4 r^5 [\frac{1}{2v^2 u^{\frac{3}{2}}} + \frac{1}{2v^2(\varepsilon r)^3} + \frac{2}{v^3 u^{\frac{1}{2}}} - \frac{2}{v^3 \varepsilon r}][\frac{2J_1(\beta)\beta^2 - 4J_1(\beta) + 2J_o(\beta)\beta}{\beta^2}] dr\}$$

In above integral one can use the following general integral:

$$\int_0^\tau ds \int_0^\infty \frac{r}{2u^{\frac{3}{2}}} A(sr) dr = \int_0^\infty \frac{\tau J_o(\beta)}{1 + r^2 + r\sqrt{1+r^2}} dr = 1 - e^{-s} \qquad (B4)$$

Finally, (20a) is found as follows:

$$(20a) = \frac{\overline{v}_1}{\overline{R}} \sigma_f^2 \lambda \{1 - e^{-\tau} - \varepsilon \int_0^\infty [2rJ_1(\beta) \frac{2u^{\frac{3}{2}} - \varepsilon r(v+2u)}{v^2 u^{\frac{3}{2}}} + F_1(r)] dr \qquad (B8)$$

where

$$F_1(r) = [\frac{(2-\beta^2)J_1(\beta) - \beta J_o(\beta)}{r\tau^2}][\frac{\varepsilon^3 r^3(v+4u) + u^{\frac{3}{2}}(5v-4u)}{v^3 u^{\frac{3}{2}}}],$$

$$u = 1 + r^2,$$

$$\beta = r\tau,$$

$$v = 1 + r^2 - \varepsilon^2 r^2,$$

**Appendix C. Derivation of equation (20c)**

Equation (20c) contains the following integral:

$$(20c) = \frac{2\bar{v}_1}{R^3} \int_0^t C_{v_1 R}(\frac{\bar{v}_1}{R}s,0,0)\, ds$$

Here we use equation (17), the cross spectrum between $v_1$ and $R$, in order to find $C_{v_1 R}(\xi)$:

$$C_{v_1 R}(\xi) = \frac{\rho_b}{n^2} K^G K_d^G J a \frac{\sinh(\sigma_w)}{\sigma_w} \frac{\sigma_f^2 \mu_1 \mu_2 \mu_3}{\pi^2} \int_{-\infty}^{\infty}\int_{-\infty}^{\infty}\int_{-\infty}^{\infty} (1-(\frac{k_1}{k})^2)^2 \frac{1}{(1+(\mu k)^2)^2} \cos(k.\xi)\, dk$$

Therefore,

$$C_{v_1 R}(\frac{\bar{v}_1}{R}s,0,0) = \frac{\rho_b}{n^2} K^G K_d^G J a \frac{\sinh(\sigma_w)}{\sigma_w} \frac{\sigma_f^2 \mu_1 \mu_2 \mu_3}{\pi^2} \int_{-\infty}^{\infty}\int_{-\infty}^{\infty}\int_{-\infty}^{\infty} (1-(\frac{k_1}{k})^2)^2 \frac{1}{(1+(\mu k)^2)^2} \cos(k_1 \frac{\bar{v}_1}{R}s)\, dk$$

$$\underset{k_i' = \mu_i k_i}{=} \frac{\rho_b}{n^2} K^G K_d^G J a \frac{\sinh(\sigma_w)}{\sigma_w} \frac{\sigma_f^2 \mu_1 \mu_2 \mu_3}{\pi^2} \int_{-\infty}^{\infty}\int_{-\infty}^{\infty}\int_{-\infty}^{\infty} (\frac{\mu_2^{-2} k_2^2 + \mu_3^{-2} k_3^2}{\mu_1^{-2} k_1^2 + \mu_2^{-2} k_2^2 + \mu_3^{-2} k_3^2})^2 \frac{1}{(1+k^2)^2} \cos(k_1 \frac{\bar{v}_1}{\mu_1 R}s)\, dk$$

We change the variable as $s' = \mu_1^{-1} \bar{R}^{-1} \bar{v}_1 s$ and $\tau = \bar{v}_1 t / \mu_1 \bar{R}$ and (20 c) changes to:

$$(20c) = \frac{2}{\bar{R}^2} \frac{\rho_b}{n^2} K^G K_d^G J a \frac{\sinh(\sigma_w)}{\sigma_w} \frac{\sigma_f^2 \mu_1}{\pi^2} \int_0^\tau \int_{-\infty}^{\infty}\int_{-\infty}^{\infty}\int_{-\infty}^{\infty} (\frac{\mu_2^{-2} k_2^2 + \mu_3^{-2} k_3^2}{\mu_1^{-2} k_1^2 + \mu_2^{-2} k_2^2 + \mu_3^{-2} k_3^2})^2 \frac{1}{(1+k^2)^2} \cos(k_1 s')\, dk ds$$

Now we have $\mu_1 = \mu_2 = \lambda$ and $\mu_3 = \varepsilon\lambda$. Thus

$$(20c) = \frac{2}{\bar{R}^2} \frac{\rho_b}{n^2} K^G K_d^G J a \frac{\sinh(\sigma_w)}{\sigma_w} \frac{\sigma_f^2 \mu_1}{\pi^2} \int_0^\tau \int_{-\infty}^{\infty}\int_{-\infty}^{\infty}\int_{-\infty}^{\infty} (1 - \frac{k_1^2}{k_1^2 + k_2^2 + \varepsilon^{-2} k_3^2})^2 \frac{1}{(1+k^2)^2} \cos(k_1 s)\, dk ds$$

Now changing to spherical coordinate system and defining:

$$k_1 = r\cos\beta, \ k_2 = r\sin\beta, \ k_3 = k_3$$
$$dk_1 k_2 k_3 = r \, dr \, d\theta \, dk_3$$

Therefore we get:

$$(20c) = \frac{2}{\overline{R}^2}\frac{\rho_b}{n^2}K^G K_d^G J a \frac{\sinh(\sigma_w)}{\sigma_w}\frac{\sigma_f^2 \lambda}{\pi^2}\int_0^{\tau}ds\int_0^{\infty}r\,dr\int_0^{2\pi}d\theta\int_{-\infty}^{\infty}(1-\frac{r^2\cos^2\theta}{r^2+\varepsilon^{-2}k_3^2})\frac{\cos(sr\cos\theta)}{(1+r^2+k_3^2)}dk_3 \quad (C1)$$

In order to derive the integral in equation (C1) we use the general integrals in (B2a) and (B2b). Furthermore, we use (B3a) and (B3b) along with (B4). The final form of equation (20c) is presented as follows:

$$(20c) = \frac{2}{\overline{R}^2}\frac{\rho_b}{n^2}K^G K_d^G J a \frac{\sinh(\sigma_w)}{\sigma_w}\sigma_f^2 \lambda \ \{1-e^{-\tau}+2\varepsilon^2\int_0^{\infty}r^2[\frac{1}{2vu^{\frac{3}{2}}}+\frac{1}{v^2 u^{\frac{1}{2}}}-\frac{1}{v^2\varepsilon r}]J_1(\beta)dr\} \quad (C2)$$

**Appendix D. Derivation of equation (20b)**

Equation (20b) contains the following integral:

$$(20b) = \frac{\overline{v_1}^2}{\overline{R}^2}\int_0^t C_{RR}(\frac{\overline{v_1}}{\overline{R}}s,0,0)ds$$

$C_{RR}(\xi)$ is as follows (see equation 12):

$$C_{RR}(\xi) = (\frac{\rho_b}{n}K_d^G)^2 e^{[\sigma_w^2]}(e^{C_{ww}(\xi)}-1)$$

where

$$C_{ww}(\xi) = \sigma_w^2 e^{(-|\frac{\xi}{\mu}|)} = \sigma_w^2 e^{(-\sqrt{(\frac{\xi_1}{\mu_1})^2 + (\frac{\xi_2}{\mu_2})^2 + (\frac{\xi_3}{\mu_3})^2})}$$

We consider the mean flow direction along $x$ axis. Thus,

$$C_{ww}(\xi,0,0) = \sigma_w^2 e^{(-|\frac{\xi}{\mu}|)} = \sigma_w^2 e^{(-\frac{\xi_1}{\mu_1})}$$

We choose $\mu_1 = \lambda$ so that we can calculate (20b) as:

$$(20b) = \frac{\bar{v}_1^2}{R^2} \int_0^t C_{RR}(\frac{\bar{v}_1}{R} s, 0, 0) ds = \frac{\bar{v}_1^2}{R^2} (\frac{\rho_b}{n} K_d^G)^2 e^{[\sigma_w^2]} \int_0^t (e^{\sigma_w^2 e^{(-\frac{\bar{v}_1}{R\lambda}s)}} - 1) ds \qquad (D1)$$

Integral in (D1) can be easily derived as explained below. In terms of multimodal porous media the derivation is not straight forward and a numerical integration is performed. In order to derive (D1) for unimodal porous media we use the following singular integrals.

$$Ei(x) = -\int_{-x}^{\infty} \frac{e^{-s}}{s} ds \qquad (D2)$$

(D2) is known as exponential integral.

$$Ein(x) = -\int_0^{-x} \frac{1 - e^{-s}}{s} ds = \gamma + \ln(x) + Ei(x) = -\sum_{i=1}^{n} \frac{x^n}{n\, n!} \quad \text{for} \quad x > 0 \qquad (D3)$$

(D3) is known as Entire function, and $\gamma = 0.577...$ is the Euler's constant. Using the above integrals we can derive the following integral:

$$\int_0^t e^{Ae^{-\alpha\tau}} d\tau = \sum_{n=0}^{\infty} \int_0^t \frac{A^n e^{-n\alpha\tau}}{n!} d\tau = \sum_{n=1}^{\infty} \frac{1}{-n\alpha} \frac{A^n}{n!}(e^{-n\alpha t}-1) + t$$

$$= -\frac{1}{\alpha}\sum_{n=1}^{\infty} \frac{A^n e^{-n\alpha t}}{nn!} + \frac{1}{\alpha}\sum_{n=1}^{\infty} \frac{A^n}{nn!} + t$$

$$= \frac{1}{\alpha}[h(A) - h(Ae^{-\alpha t})] + t$$

$$= \frac{1}{\alpha}[Ei(A) - Ei(Ae^{-\alpha t})] \qquad (D4)$$

Therefore,

$$\int_0^t e^{Ae^{-\alpha\tau}} d\tau = \frac{1}{\alpha}[Ei(A) - Ei(Ae^{-\alpha t})] \qquad (D5)$$

Therefore, in case of unimodal distribution of $C_{ww}(\xi)$ the result of (20b) is:

$$(20b) = \frac{\overline{v}_1}{\overline{R}}(\frac{\rho_b}{n} K_d^G)^2 e^{[\sigma_w^2]} \lambda [Ei(\sigma_w^2) - Ei(\sigma_w^2 e^{-\frac{\overline{v}}{\overline{R}\lambda}t})] \qquad (D5)$$